\documentclass[doublecol]{epl2}

\usepackage{setspace}

\usepackage{float} 

\usepackage{amsmath}
\usepackage{hyperref}
\usepackage{graphicx}
\usepackage{amsfonts}
\usepackage{amsthm}
\usepackage{cases}
\usepackage{bm}

\usepackage{fixmath}

\usepackage{enumitem} 

\usepackage[normalem]{ulem}

\usepackage{color}
\definecolor{Blue}{rgb}{0.00, 0.00, 0.80}
\definecolor{Red}{rgb}{0.80, 0.00, 0.00}
\definecolor{Green}{rgb}{0.00, 0.50, 0.00}

\hypersetup{
    colorlinks=true,       
    linkcolor=red,          
    citecolor=blue,        
    filecolor=magenta,      
    urlcolor=cyan           
}

\newcommand{\nn}{\nonumber}
\newcommand{\be}{\begin{equation}}
\newcommand{\ee}{\end{equation}}
\newcommand{\bea}{\begin{eqnarray}}
\newcommand{\eea}{\end{eqnarray}}

\newcommand{\beq}{\begin{equation}}
\newcommand{\eeq}{\end{equation}}
\newcommand{\beqn}{\begin{eqnarray}}
\newcommand{\eeqn}{\end{eqnarray}}

\begin{document}

\title{Striking universalities in stochastic resetting processes}

\author{Naftali R. Smith\inst{1} \and Satya N. Majumdar\inst{2} \and Gr\'egory Schehr\inst{3}}
\shortauthor{N. R. Smith \etal}

\institute{                    
  \inst{1} Department of Solar Energy and Environmental Physics, Blaustein Institutes for Desert Research, Ben-Gurion University of the Negev, Sede Boqer Campus, 8499000, Israel\\
  \inst{2} Universit{\'e} Paris-Saclay, CNRS, LPTMS, 91405, Orsay, France\\
  \inst{3} Sorbonne Universit{\'e}, Laboratoire de Physique Th{\'e}orique et Hautes Energies, CNRS UMR 7589, 4 Place Jussieu, 75252 Paris Cedex 05, France
}

%

\date{\today}

\abstract{
Given a random process $x(\tau)$ which undergoes stochastic resetting at a constant rate $r$ to a position drawn from a distribution ${\cal P}(x)$, we consider a sequence of dynamical observables $A_1, \dots, A_n$ associated to the intervals between resetting events. We calculate exactly the probabilities of various events related to this sequence: that the last element is larger than all previous ones, that the sequence is monotonically increasing, etc. Remarkably, we find that these probabilities are ``super-universal'', i.e., that they are independent of the particular process $x(\tau)$, the observables $A_k$'s in question and also the resetting distribution ${\cal P}(x)$. For some of the events in question, the universality is valid  provided certain mild assumptions on the process and observables hold (e.g., mirror symmetry).
}

\maketitle



One of the main goals of statistical mechanics is to find universal laws that describe the behavior of broad classes of systems. In this paper, we discover such laws for a class of nonequilibrium stochastic systems that has attracted much interest, especially over the last decade: stochastic processes with random resetting to some state (which is usually the initial state), see \cite{EMS20,GJ22,PKR22} for recent reviews. Consider for example the simplest setting where a single Brownian particle diffuses with a diffusion constant $D$ and resets to its initial position, say the origin, with a constant rate $r$ \cite{EM1,EM2}. There are two interesting consequences of resetting: (i) the resetting drives the system into a non-equilibrium stationary state where the distribution of the position of the particle becomes independent of time and is typically non-Gaussian, (ii) the mean first-passage time (MFPT) to a target located at a distance $R$ from the origin becomes finite and, moreover, as a function of the resetting rate $r$, the MFPT exhibits a minimum indicating the existence of an optimal resetting rate $r^*$ \cite{EM1,EM2,EM14}. These two features have been found in numerous theoretical models, going beyond simple diffusion: random walk on a lattice with resetting \cite{BMM22}, continuous-time random walks~\cite{MV13,MC16,MMV17,MM19} and L\'evy flights with resetting \cite{KMSS14,KG19}, Brownian particle in a confining potential \cite{Pal15}, active run-and-tumble particles under resetting \cite{EM18,Mas19,TGMS22}, non-Poissonian resetting \cite{PKE16,NG16}, Poissonian resetting with a site-dependent resetting rate \cite{Pin20,BMG22}, resetting with memory \cite{BS14,BEM17}, etc. Moreover, these two features have been verified in recent experiments using optical tweezers, both in one~\cite{FPSRR20, BBPMC20} and two dimensions~\cite{FBPCM21}. The relaxation dynamics to the stationary state exhibits a phase transition in the associated large deviation function~\cite{MSS15a}. The conditions for the existence of an optimal resetting rate for search processes have been studied in various contexts \cite{Reuveni16,PR17,Pal19,PKR22}. Besides, recent years have seen a growing number of applications of stochastic resetting in various areas~\cite{RLSTG16,CS18,DH2019,Bres20,BRR20,SW21,SSIM21,SM22,SGS22,VCWMS22,DM23}.

\begin{figure}
\includegraphics[width=0.8\linewidth]{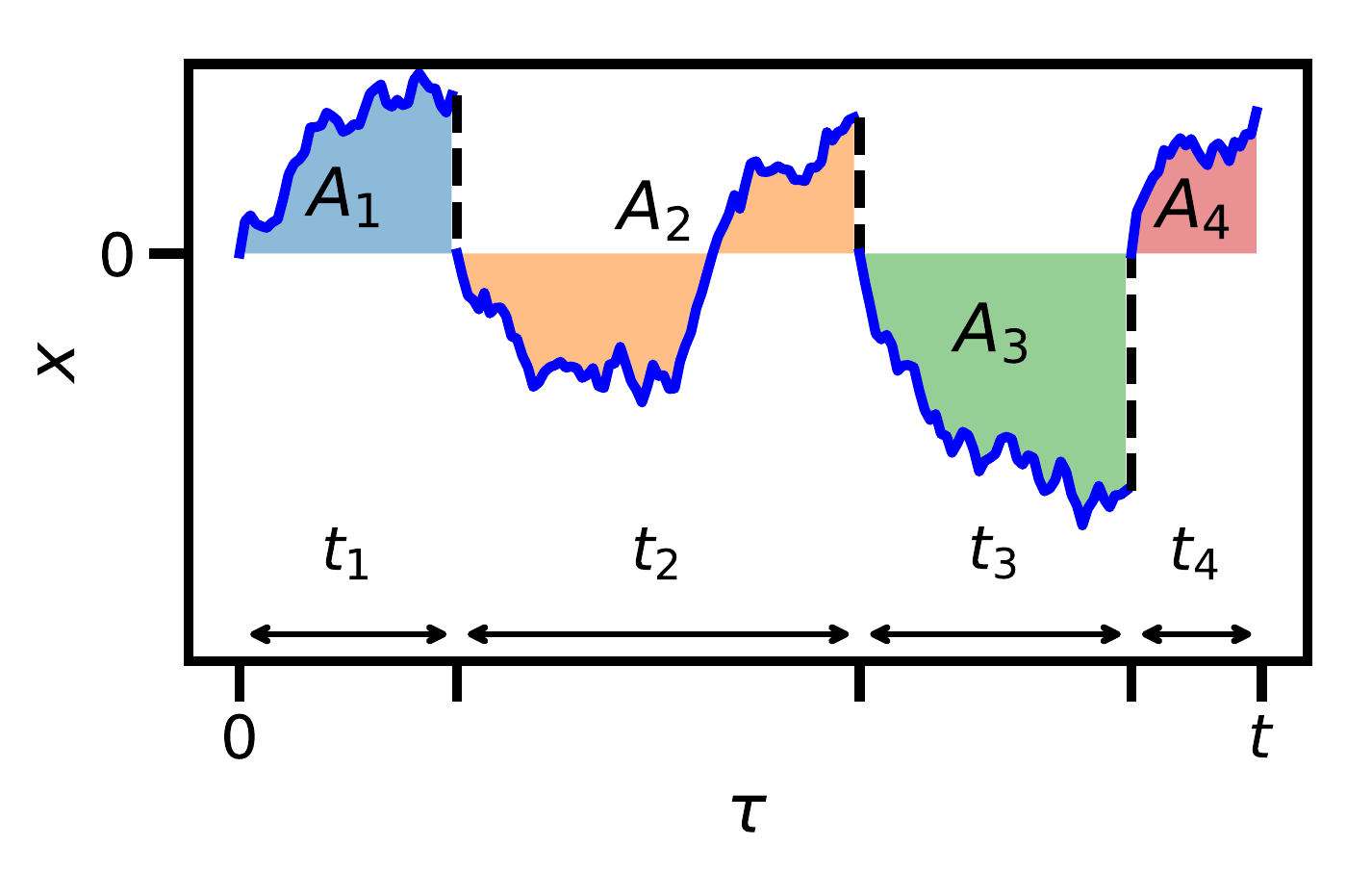}
\caption{Schematic plot of a random process $x(\tau)$ with stochastic resetting to the origin, corresponding to $\mathcal{P}(x) = \delta(x)$. We are interested in statistical properties that concern dynamical observables $A_1, \dots, A_n$  associated to the intervals between resettings, see eq.~\eqref{Akdef}. The figure represents the particular case $U(z)=z$, for which $A_k$'s are the areas under $x(\tau)$ in each interval.
The vertical dashed lines represent the resetting events that are separated by the time intervals $t_1,t_2, t_3,\dots$. }
\label{FigSchematicRBM}
\end{figure}

The goal of this Letter is to present a class of universal probabilities associated with the resetting of any stochastic processes, not necessarily simple diffusion. Consider any stochastic process $x(\tau)$ (not necessarily Markovian) whose initial condition $x_0 = x(0)$ is sampled from some given probability distribution function (PDF) $\mathcal{P}(x_0)$, and evolves according to its own noisy dynamics, but, with a constant rate $r$ (Poissonian resetting), is stochastically reset to a new position that is again sampled from the distribution $\mathcal{P}(x)$. The process evolves up to a total time $t$. For a given realization of the process, we denote the durations of the successive intervals by $t_1, \dots, t_n$. Here $n$ denotes the number of intervals up to $t$, or equivalently $(n-1)$ denotes the number of resettings up to time $t$. For fixed $t$, clearly $n$ ($n=1,2,\dots$) is a random variable that varies from one realization of the process to another. Note that the last interval of the process is yet to reset, see fig.~\ref{FigSchematicRBM}.
We also denote by $0 = \tau_1 < \tau_2 < \dots < \tau_n < \tau_{n+1} = t$ the endpoints of the intervals, such that $t_k = \tau_{k+1} - \tau_k$.  In the Poissonian protocol, the interval $t_k$ between the $(k-1)^{\rm th}$ and $k^{\rm th}$ resetting is distributed according to an exponential distribution $p(t_k) = r\, e^{-r\,t_k}$.

Associated with each interval of duration $t_k$, we define an observable
\be
\label{Akdef}
A_{k}=\int_{\tau_{k}}^{\tau_{k+1}}U\left(x\left(\tau\right)\right)d\tau\,,
\ee
where $U(z)$ can be any function. For example, if $U(z) = 1$, then $A_k = t_k$ is just the duration of the interval between the $(k-1)^{\text th}$ and the $k^{\rm th}$ resetting event. Similarly if $U(z) = z$, then $A_k$ is the area under the process between the $(k-1)^{\text{th}}$ and $k^{\text{th}}$ resetting.
Thus we have a sequence of random variables $\left\{ A_{1},\dots,A_{n}\right\} $ associated to any realization of the process.
For convenience, even for general $U(z)$, we will refer to $A_k$ as the ``area'' associated to the $k^{\rm th}$ interval. Even though we are considering here the resetting process, these sequences can also be defined for arbitrary renewal processes.
It is then natural to ask different questions concerning the random variables $A_1, \dots, A_n$. 
For instance, what is the probability
\footnote{In eq.~\eqref{def_Q1}, for the case $n=1$ the probability is to be understood to equal $1$ (i.e., in this case, there is just one interval so $A_1$ is defined to be the largest of the $A_i$'s).}
\be \label{def_Q1}
Q_{1}\left(t\right)=\sum_{n=1}^{\infty}\text{Prob}\left(A_{n}>\max\left\{ A_{1},\dots,A_{n-1}\right\} \,|\,t\right)
\ee
that the area of the last interval is bigger than all the previous ones?
 For $U(z)=1$, for instance, $Q_1(t)$ represents the probability that the last interval is the longest. Indeed, this probability was studied in the context of returns to the origin of random walks and L\'evy flights in one dimension \cite{GMS09}. The same question, but with $U(z) = |z|$,  arises quite naturally in another well studied stochastic process, as we show now. This concerns the dynamics of  
 a run-and-tumble particle (RTP) in one dimension \cite{Kac74,Weiss02,TC08,CT15}. The RTP represents the dynamics of an active particle whose motion alternates between ballistic runs of random durations and instantaneous tumblings. The duration $t_k$ of the $(k-1)^{\rm th}$ run is chosen from an exponential distribution $p(t_k) = r\, e^{-r\,t_k}$ where $1/r$ represents the persistence time and the constant velocity $v$ during this run is chosen from an arbitrary distribution ${W}(v)$. At the end of each run, the particle tumbles instantaneously, i.e., chooses a new velocity, again from the distribution ${W}(v)$ and starts a new run of random duration \cite{MoriPRL20,MoriPRE20}. 
Here we choose the relevant stochastic process $x(\tau)$ in eq.~\eqref{Akdef} to be the velocity $v(\tau) =v$ between tumblings, $U(z) = |z|$ and ${\cal P}(x) \equiv W(v)$. 
Then the area 
$A_k= \int_{\tau_k}^{\tau_{k+1}} |{v}| d\tau= |{v}| t_k$ represents the run length between the $(k-1)^{\rm th}$ and the $k^{\rm th}$ tumblings.
In this context, $Q_1(t)$ is just the probability that the length of the last run before $t$ is the longest one.

\begin{figure}
\includegraphics[width=0.8\linewidth]{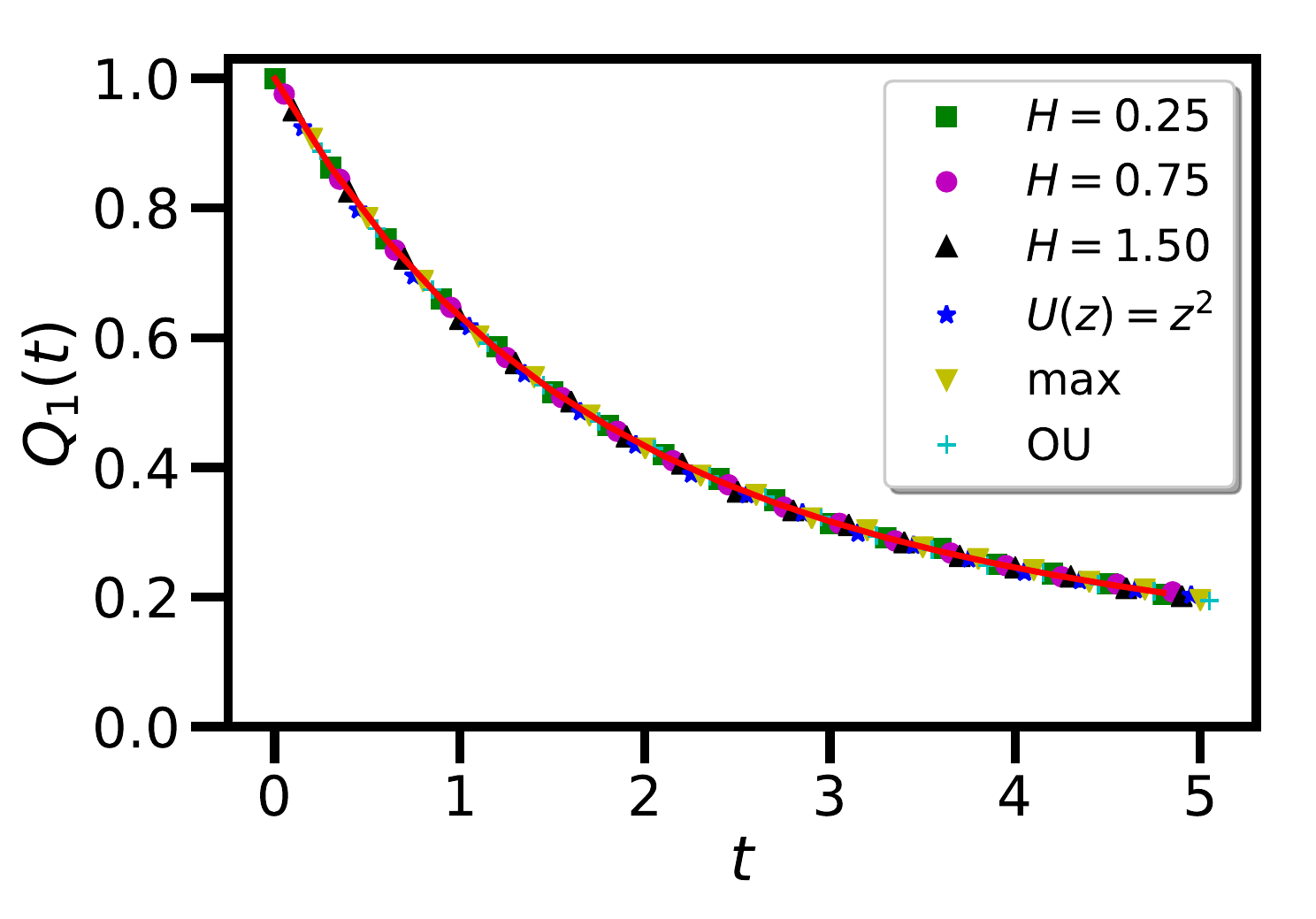}
\caption{Probability $Q_1(t)$ that $A_{n}>\max\left\{ A_{1},\dots,A_{n-1}\right\} $, as a function of time $t$. The solid line represents our exact, universal result in eq.~\eqref{Q1sol}. 
Symbols represent simulations of different random processes and observables, as indicated in the figure legend. With the choice $U(z)=z$, we have measured $Q_1(t)$ for different processes such as the fractional Brownian motion with different choices of the Hurst exponent $H$, the random acceleration process, and the Ornstein-Uhlenbeck process (OU). Furthermore, with the choices $U(z)=z^2$ and  
$A_{k}=\underset{\tau_{k} \leq \tau \leq \tau_{k+1}}{\max}x\left(\tau\right)$, we have considered the standard Brownian motion.
We have chosen $r=1$ and $\mathcal{P}\left(x\right)=\delta\left(x\right)$ in all examples, except for the OU case where ${\cal P}(x)$ is chosen to be the stationary reset-free distribution of the OU process (Gaussian).}
\label{FigQHs1}
\end{figure}

 In both examples above, the underlying stochastic process can be thought of as special cases of a resetting process $x(\tau)$.   
Thus studying this probability $Q_1(t)$ is physically relevant in different contexts and it is natural to study this for a generic process $x(\tau)$ with Poissonian resetting, with an arbitrary choice of $U(z)$ in eq.~\eqref{Akdef} and an arbitrary resetting PDF ${\cal P}(x)$. Naively, one would expect this probability $Q_1(t)$ to depend on the process $x(\tau)$, on $U(z)$ and on $\mathcal{P}(x)$.
However, performing simulations for different processes $x(\tau)$, with different functions $U(z)$ and PDFs $\mathcal{P}(x)$
(see fig.~\ref{FigQHs1}) we found, to our great surprise, that $Q_1(t)$ is seemingly independent of $x(\tau)$, $U(z)$ and $\mathcal{P}(x)$!

The principal goal of this Letter is to uncover the mechanism behind this ``super-universality''. We show that, for resetting processes with Poissonian protocol, the probability $Q_1(t)$ is indeed independent of the process $x(\tau)$, the function $U(z)$ and the resetting point distribution ${\cal P}(x)$ for any $t$ (and not just for large $t$),  as long as the $A_k$'s are continuous random variables (which will be assumed henceforth) and is given by a remarkably simple and exact formula
\be
\label{Q1sol}
Q_{1}\left(t\right)=F_{1}\left(rt\right),\quad F_{1}\left(z\right)=\frac{1-e^{-z}}{z}\,.
\ee
In fig.~\ref{FigQHs1}, we compare our formula for $Q_1(t)$ with numerical simulations for different processes $x(\tau)$ -- such as the fractional Brownian motion, the random acceleration process and the Ornstein-Uhlenbeck processes -- with different choices of $U(z)$ and ${\cal P}(x)$. We see that all the numerical data fall on top of the predicted universal formula in eq.~\eqref{Q1sol}.  
Furthermore, we will see that a similar ``super-universality'' also holds for various other observables, going beyond $Q_1(t)$. 
The common feature of these observables is that they all involve the probability of an event related to the ordering 
of the $A_k$'s. For example, we study the probability that the set $A_1 ,\dots, A_n$ is monotonically increasing (or decreasing), i.e.,
\be \label{defQ2}
Q_{2}\left(t\right)=\sum_{n=1}^{\infty}\text{Prob}\left(A_{1}<A_{2}<\dots<A_{n}\,|\,t\right) \;.
\ee
In some special cases of the process $x(\tau)$, this probability $Q_2(t)$ appeared quite naturally, e.g., in the context of the statistics of records, 
with applications to earthquakes dynamics and finance \cite{MBN13, GMS16}. For Poissonian resetting processes,   
we show that $Q_2(t)$ is again independent of the process $x(\tau)$, the choice of $U(z)$ and ${\cal P}(x)$ and is given by the universal formula  
\be 
\label{res_Q2}
Q_{2}  \left(t\right) = F_{2} \left(rt\right),\quad \text{with}\quad F_2 \left(z\right) = \frac{e^{-z}}{\sqrt{z}}\, I_1(2\,\sqrt{z})\,,
\ee
where $I_k(z)$ is the $k^{\rm th}$ modified Bessel function of the first kind. The function $F_2(z) \to 1$ as $z \to 0$ and decays as 
$F_2(z) \simeq z^{-3/4}\,e^{2\sqrt{z}-z}/(2 \sqrt{\pi})$ as $z \to \infty$. This universal result \eqref{res_Q2} is also verified in numerical simulations, see fig.~S4 in the Supplementary Material (SM) \cite{SM}. Later, we provide several other examples of such super-universal probabilities.

We start by briefly outlining the derivation of the universal result in eq.~\eqref{Q1sol}. We consider a trajectory of the process $x(\tau)$  starting at $x_0$ drawn from the resetting distribution ${\cal P}(x_0)$ at $\tau=0$, see fig.~\ref{FigSchematicRBM}. The configuration of this trajectory is specified by the following set
of random variables: 
(i) the sequence $\{A_1, A_2, \cdots, A_n\}$ denoting the successive areas between resettings, 
(ii) the durations $\{t_1, t_2, \cdots, t_n\}$ of the successive intervals~(see fig.~\ref{FigSchematicRBM}) and 
(iii) the number of intervals $n$ up to time $t$. 
Under resetting, the random variables $A_k$'s  (conditioned on the $t_k$'s) are statistically independent and we assume that each is distributed via the PDF $p\left(A_{k}\,|\,t_{k}\right)$, normalised to unity $\int_{-\infty}^{\infty}p\left(A_k\,|\,t_k\right)dA_k = 1$.  Of course $p\left(A_{k}\,|\,t_{k}\right)$ depends on the details of the process $x(\tau)$, on the function $U(z)$ and on the distribution $\mathcal{P}(x)$. Here, we consider $p\left(A_{k}\,|\,t_{k}\right)$ to be given and its detailed form will be of no consequence, as we will see.

It is useful to start with the joint distribution $P\left[\left\{ A_{k}\right\} ,\left\{ t_{k}\right\} ,n\,|\,t\right]$ of the variables (i)-(iii) above.  
If the interval distribution is denoted by $p(t_k)$, then this joint distribution can be expressed as
\bea
\label{jointAtaun0}
\!\!\!\!  P \! \left[\left\{ A_{k}\right\} ,\left\{ t_{k}\right\} ,n | t\right] \! &=&  \left[\prod_{k=1}^{n-1}p\left(t_{k}\right)p\left(A_{k}\,|\,t_{k}\right)\right] \nonumber \\
&\times&  q\left(t_{n}\right)p\left(A_{n}\,|\,t_{n}\right)\delta\!\left(\sum_{k=1}^{n}t_{k}-t\!\right)\!,
\eea  
where $q\left(t_{n}\right)=\int_{t_{n}}^{\infty}p\left(t'\right)\,dt'$.
Note that the statistical weight $q\left(t_{n}\right) p\left(A_{n}\,|\,t_{n}\right)$ associated to the last interval is thus different from the preceding ones. This is because the last interval remains incomplete (i.e., not yet reset) at the final time $t$. The delta function in \eqref{jointAtaun0} ensures that the total time is $t$ and its presence makes the intervals correlated. For the Poissonian resetting $p(t_k) = r\,e^{-r t_k}$,  however, one has $q\left(t_{n}\right)=e^{-rt_{n}}$, which is of the same functional form as the $p(t_k)$'s up to the multiplicative constant $r^{-1}$. Therefore, in this particular case, eq.~(\ref{jointAtaun0}) simplifies to  a form that is manifestly symmetric with respect to exchanging any two of the intervals (this may include the last interval),
\be \label{joint_exp}
P \! \left[\left\{ A_{k}\right\} ,\left\{ t_{k}\right\} ,n | t\right] \! = \! \frac{1}{r} \!  \left[\prod_{k=1}^{n}\! re^{-rt_{k}}p\left(A_{k}|t_{k}\right)\right] \!  \delta \! \left(\sum_{k=1}^{n} \! t_{k}\!-t\! \right) \! .
\ee
In terms of the joint distribution, the probability $Q_1(t)$ is given by
\be
\label{Q1t1_txt}
Q_{1} \! \left(t\right) \! = \! \sum_{n=1}^{\infty} \! \int \!  d\vec{A} \! \int  \! d\vec{t} \, P\left[\left\{ A_{k}\right\} ,\left\{ t_{k}\right\} ,n\,|\,t\right]\prod_{k=1}^{n-1} \! \theta\left(A_{n}-A_{k}\right),
\ee
where we introduced shorthand notations $\int d\vec{A}\equiv\int_{-\infty}^{\infty}dA_{1}\dots\int_{-\infty}^{\infty}dA_{n}$ and $\int d\vec{t}=\int_{0}^{\infty}dt_{1}\dots\int_{0}^{\infty}dt_{n}$.
We now plug \eqref{joint_exp} into \eqref{Q1t1_txt} and take a Laplace transform with respect to time $t$. This decouples the different intervals and one gets for the Laplace transform of $Q_1(t)$
\be
\label{Q1s1}
\! \tilde{Q}_{1}\left(s\right)=\!\! \int_{0}^{\infty}Q_{1}\left(t\right)e^{-st}dt=\frac{1}{r}\sum_{n=1}^{\infty}\left(\frac{r}{r+s}\right)^{n} \! p_{n}\!\left(s\right) ,
\ee
where $p_n(s)$ is given by
\be
\label{pns}
p_{n}\left(s\right)=\int d\vec{A} \, \prod_{i=1}^{n}\phi_{s}\left(A_{i}\right)\prod_{k=1}^{n-1}\theta\left(A_{n}-A_{k}\right),
\ee
and
\be
\label{phisA}
\phi_{s}\left(A\right)=\left(r+s\right)\int_{0}^{\infty}e^{-\left(r+s\right)t'}p\left(A\,|\,t'\right)dt' \, .
\ee
The function $\phi_s\left(A\right)$ is nonnegative and normalised to unity $\int_{-\infty}^{\infty}\phi_{s}\left(A\right)dA=1$. Hence it can be interpreted as an auxiliary PDF parametrised by $s$.  In turn, $p_n(s)$ can be interpreted as the probability that, for a set $A_1, \dots, A_n$ of $n$ independent and identically distributed (i.i.d.) random variables each distributed via the PDF $\phi_s(A)$, the last entry $A_n$ is bigger than all the previous ones (i.e., it is maximal). Note that the dependences on the process $x(\tau)$, the function $U(z)$ and the resetting PDF ${\cal P}(x)$ are all contained in $\phi_s(A)$. Due to the symmetry under the exchange $A_{i}\leftrightarrow A_{j}$, this probability is clearly $p_n(s) = 1/n$, independent of $\phi_s(A)$ [and hence of the process  $x(\tau)$, the function $U(z)$ and ${\cal P}(x)$]. Substituting this result in \eqref{Q1s1}, we thus obtain the universal result
\be
\label{Q1s2}
\tilde{Q}_{1}\left(s\right)=\frac{1}{r}\sum_{n=1}^{\infty} \frac{1}{n} \left(\frac{r}{r+s}\right)^{n}=\frac{1}{r}\ln\left(\frac{s+r}{s}\right) \, .
\ee
Inverting this Laplace transform gives the explicit result in eq.~\eqref{Q1sol}.

We find that, remarkably, this universality extends to a much broader class of questions that involve ordering and/or sign properties of $A_1, \dots, A_n$, with several important particular cases solved explicitly below.
A general theoretical framework that solves this wider class of problems proceeds in a similar way as the example discussed above.
Let $Q(t)$ be the probability of some event that involves the sequence $A_1, \dots, A_n$, and let $\chi\left(A_{1},\dots,A_{n}\right)$ denote the indicator function of the event in question, i.e., $\chi=1$ if this event holds and $\chi = 0$ otherwise (we suppress the $n$-dependence of $\chi$ for brevity). Following a similar argument as the specific example considered above, it is clear that the Laplace transform $\tilde Q(s)$ of $Q(t)$  
still reads 
\be
\label{Qs}
\tilde{Q}  \left(s\right) =  \int_{0}^{\infty}\! Q \left(t\right) e^{-st}dt = \frac{1}{r}\sum_{n=1}^{\infty} \left( \frac{r}{r+s} \right)^{n} p_{n} \left(s\right) ,
\ee
where $p_n(s)$ is given by
\be \label{pn_gen}
p_{n}  \left(s\right)=\int d\vec{A}\,\prod_{k=1}^{n}   \phi_{s} \left(A_{k}\right)\chi\left(A_{1},\dots,A_{n}\right) \,,
\ee
with $\phi_s(A)$ in (\ref{phisA}). If the event is such that $p_n(s)=p_n$ is independent of $\phi_s(A)$ and hence of $s$ also, one can invert the Laplace transform in eq.~\eqref{Qs} explicitly
\footnote{In the Laplace inversion of eq.~\eqref{Qs}, we use
$\mathcal{L}_{s+t}^{-1}\left[\frac{1}{\left(r+s\right)^{n}}\right]=e^{-rt}\frac{t^{n-1}}{\Gamma\left(n\right)}$,
where $\mathcal{L}$ denotes the Laplace transform, yielding eq.~\eqref{QandFgeneral}.}
to get, for any $t$,  
\be
\label{QandFgeneral}
\!\!\! Q\left(t\right)=F\left(rt\right),\quad\text{where}\quad F\left(z\right)=e^{-z}\sum_{n=1}^{\infty}\frac{z^{n-1}p_{n}}{\left(n-1\right)!} \,.
\ee
Since it is independent of $\phi_s(A)$, it is clearly universal, i.e., independent of the process $x(\tau)$, the function $U(z)$ and ${\cal P}(x)$. In the  example $Q_1(t)$ discussed before, we have $p_n = 1/n$, for which eq.~\eqref{QandFgeneral} reproduces eq.~\eqref{Q1sol}. Below, we consider few other examples sharing this universal property.

We next consider the example $Q_2(t)$ defined in eq. (\ref{defQ2}). In this case, $p_n(s)$  reads
\be 
\label{pn_Q2}
p_{n}\left(s\right)=\int d\vec{A}\,\prod_{k=1}^{n}\phi_{s}\left(A_{k}\right)\prod_{k=1}^{n-1}\theta\left(A_{k+1}-A_{k}\right)\;.
\ee
Thus 
$\chi(\vec{A})=\prod_{k=1}^{n}\theta\left(A_{k+1}-A_{k}\right)$
 in this example. Since the product $\prod_{k=1}^n \phi_s(A_k)$ is 
invariant under permutations of the $A_k$'s, all permutations are equally likely, including the ordered one $A_1<A_2< \cdots < A_n$. Hence it is clear that $p_n(s) = 1/n!$\,. Substituting this in the general formula (\ref{QandFgeneral}) gives the result in eq.~\eqref{res_Q2}, upon using the series expansion of $I_1(z)$.

Another example concerns the ``occupation time probability'' that frequently appears in many stochastic processes including random walks \cite{Feller}. In our context, this corresponds to the probability that exactly $m$ of the $n$ areas $A_k$'s are positive, i.e., 
$Q^{(m)}_{3}\left(t\right)=\sum_{n=1}^{\infty}\text{Prob}\left(\sum_{k=1}^{n}\theta\left(A_{k}\right)=m\right)$.
Let us assume that and $\phi_s(A) = \phi_s(-A)$, which is indeed the case for several stochastic processes $x(\tau)$ and functions $U(z)$  (e.g., if the resetting distribution and the process are statistically invariant under sign inversion $x \to -x$, and the function $U$ is odd, $U(z) =-U(-z)$).
Then it follows that $p_n(s)$ is again independent of $\phi_s(A)$ and is simply given by the Bernoulli expression $p_n(s)=p_n = 2^{-n}{n \choose m}$. Substituting this $p_n$ in the general expression (\ref{QandFgeneral}) and carrying out the sum over $n$ explicitly (see \cite{SM}), we get $Q_{3}^{(m)}\left(t\right)=F_{3}^{(m)}\left(r\,t\right)$ where
\be
\label{F3sol}
F^{(m)}_{3}\left(z\right)=\frac{(2m+z)}{4 \, (m!) }\left(\frac{z}{2}\right)^{m-1} \! e^{-z/2} \;
\ee
is again universal for each $m \! \ge \! 0$. 
Let us briefly discuss two other examples (for details see \cite{SM}). The first concerns the probability $Q_{4}\left(t\right)= \! \sum_{n=1}^{\infty}\text{Prob}\left(A_{1}+A_{2}>0,\dots,A_{n-1}+A_{n}>0\,|\,t\right)$. This question also appears in many different contexts, notably in the study of the persistence of a stationary non-Markov sequence \cite{MD01}, where it was shown to be related to the random-field Ising model in one-dimension studied by Derrida and Gardner \cite{DG86}. In this case, it was shown that $p_n(s)$ is remarkably independent of $\phi_s(A)$ as long as it is symmetric  [$\phi_s(A) = \phi_s(-A)$] and its generating function is \cite{MD01}
\be
\label{ptildez4}
\sum_{n=0}^\infty p_n \,z^n =\frac{2}{\cot\left(z/2\right)-1}=\frac{1+\sin z-\cos z}{\cos z} \, .
\ee
Using this result, we show in \cite{SM} that $Q_4(t) = F_4(r\,t)$ where the universal scaling function $F_4(z)$ is given by
\be
\label{F4sol}
F_{4}\left(z\right)=\frac{8}{\pi^{2}}\sum_{m=-\infty}^{\infty}\frac{1}{\left(4m+1\right)^{2}}e^{-\left(1-\frac{2}{\left(4m+1\right)\pi}\right)z} \, .
\ee
The function $F_4(z) \to 1$ as $z \to 0$ and decays as 
$F_4(z) \simeq {8}\,e^{-\left(1-\frac{2}{\pi}\right)z}/\pi^2$ as $z \to \infty$.
In \cite{SM}, we verified this universal theoretical prediction in numerical simulations.

We conclude with a last example where we ask: what is the probability $Q_5(t)$ that the sequences $A_1, A_1+A_2, A_1+A_2+A_3, \cdots$ etc. are all positive? 
This question naturally arises in the context of random walks and RTP \cite{MoriPRL20,MoriPRE20,ACDK14}. For example, in the context of RTP, if one chooses the velocity $v(\tau)$ between tumblings as the relevant stochastic process and $U(z) = z$, then $A_k$'s are just the jumps in positions between two successive tumblings. As before, since $v(\tau)$ is the relevant process, one identifies the resetting distribution ${\cal P}(x)$ with $W(v)$. 
Consequently, for the RTP, $Q_5(t)$ is just the probability that the particle, starting at the origin, stays positive up to time $t$, which was studied in ref. \cite{MoriPRL20,MoriPRE20} and was found to be universal. Here, our $Q_5(t)$ applies to a more general process, where we again identify 
$S_k = A_1 + A_2 + \cdots+ A_k$ as the position of a random walker after $k$ steps, starting from $S_0=0$. The jumps $A_k = S_{k}-S_{k-1}$ are i.i.d. random variables each drawn from $\phi_s(A)$. Thus, as in the case of the RTP, the probability $p_n(s)$ in eq.~\eqref{pn_gen} is the probability that this more general random walk, starting from the origin, stays positive up to step $n$. Assuming that  $\phi_s(A) = \phi_s(-A)$, it is a remarkable fact that $p_n(s) = p_n = {2n \choose n}\,2^{-2n}$ is completely universal, i.e., independent of the jump distribution $\phi_s(A)$, thanks to the Sparre Andersen theorem~\cite{SA_54}. Substituting this $p_n$ in eq.~\eqref{QandFgeneral}, and using series expansions of $I_k(z)$, one finds $Q_5(t) = F_5(r\,t)$ where 
\be
\label{F5sol}
F_{5}\left(z\right)=\frac{1}{2}e^{-z/2}\left[I_{0}\left(\frac{z}{2}\right)+I_{1}\left(\frac{z}{2}\right)\right] \;,
\ee
where $F_5(z) \to 1/2$ as $z \to 0$ and $F_5(z) \simeq 1/\sqrt{\pi \,z}$ as $z \to \infty$. This result, found in ref. \cite{MoriPRL20,MoriPRE20} for the special case of the RTP, actually holds for more general resetting processes $x(\tau)$ and arbitrary choices of the functional $U(z)$ and ${\cal P}(x)$. These results were verified in numerical simulations (see \cite{SM}). 

To summarise, we have demonstrated a striking universal behavior of a class of probabilities $Q_i(t)$ with $i=1,\cdots, 5$ associated with generic stochastic processes that undergo Poissonian resetting. We have shown that these probabilities are universal for all $t$ (and not just for large $t$), i.e., they are independent  (for $i=3,4,5$, under the assumption of mirror symmetry) of the underlying stochastic resetting process $x(\tau)$, of the choice of the functional $U(z)$ as well as the resetting PDF ${\cal P}(x)$. Here, for simplicity, we have considered $x(\tau)$ to be a one-dimensional process but, clearly, this universality holds for higher dimensional processes $\vec{x}(\tau)$ as well, i.e., when $A_k = \int_{\tau_k}^{\tau_{k+1}} U(\vec{x}(\tau))\, d\tau$. 
In addition, this universality extends beyond the particular choice of functionals $A_k$'s in eq.~\eqref{Akdef}. For example, one can choose $A_k$ the maximum of the process $x(\tau)$ in the interval $[\tau_k,\tau_{k+1}]$ -- in that case it is not difficult to see that the probabilities $Q_i(t)$ remain the same (this is verified numerically in fig.~\ref{FigQHs1} for the Brownian motion). Another interesting question is whether this universality extends beyond the Poissonian resetting protocol, i.e., with a non-exponential $p(t_k)$. In this case, we anticipate that this universality will hold only for large $t$, provided $p(t_k)$ decays sufficiently fast  at $ t_k \to \infty$. It would be interesting to investigate these universal properties for arbitrary $p(t_k)$.

{}

\begin{widetext}
	
\begin{spacing}{1.4}

\noindent	\textbf{\large Supplemental Material to the paper {``Striking universalities in stochastic resetting processes"} by N. R. Smith, S. N. Majumdar and G. Schehr\\\\} 
\end{spacing}
\end{widetext}

\newpage

\renewcommand{\theequation}{S\arabic{equation}}
\renewcommand{\thefigure}{S\arabic{figure}}
\setcounter{equation}{0}




\noindent \textbf{Explicit expression for $F^{(m)}_3(z)$\\}

Here we derive the explicit expression of $F^{(m)}_3(z)$ in Eq. (\ref{F3sol}) in the main text. Substituting $p_n = 2^{-n} {n \choose m}$ in Eq. (\ref{QandFgeneral}) of the main text, we obtain
\bea
\label{F3m1}
F_{3}^{(m)}\left(z\right)&=&e^{-z}\sum_{n=m}^{\infty}\frac{z^{n-1}}{\left(n-1\right)!}{n \choose m}2^{-n} \nn\\
&=&e^{-z}\sum_{n=m}^{\infty}\frac{nz^{n-1}}{k!\left(n-m\right)!}2^{-n}\nn\\
&=&\left(\frac{z}{2}\right)^{m-1}\frac{e^{-z}}{4\left(m!\right)}\sum_{n=m}^{\infty}\frac{2n\left(\frac{z}{2}\right)^{n-m}}{\left(n-m\right)!}\;.
\eea
Making the change of variable $\ell = n - m$ in the last expression, we get 
\bea
 \label{F3m2}
F^{(m)}_{3}\left(z\right)&=& \left(\frac{z}{2}\right)^{m-1}\frac{e^{-z}}{4\left(m!\right)} \sum_{\ell=0}^{\infty}\frac{2\left(\ell+m\right)\left(\frac{z}{2}\right)^{\ell}}{\ell!} \nn\\
&=&\left(\frac{z}{2}\right)^{m-1}\frac{e^{-z}}{4\left(m!\right)}\left(2m+z\right)\sum_{\ell=0}^{\infty}\frac{\left(\frac{z}{2}\right)^{\ell}}{\left(\ell\right)!} \nn\\
&=&\frac{\left(2m+z\right)}{4\left(m!\right)}\left(\frac{z}{2}\right)^{m-1}e^{-z/2} \, .
\eea
This reproduces the right hand side of Eq.~\eqref{F3sol} in the main text.

\bigskip\bigskip

\noindent \textbf{Laplace inversion of $\tilde{Q}_4(s)$\\}

We start from the general expression in Eq. (\ref{Qs}) in the main text and focus on $Q_4(t)$, for which $p_n(s) = p_n$ and its generating function is given in Eq. (\ref{ptildez4}) of the main text. Using $z = r/(r+s)$, we get from Eqs. (\ref{Qs}) and (\ref{ptildez4})
\be \label{Q4_s}
\tilde Q_4(s) = \int_0^\infty Q(t)\,e^{-st}\, dt = \frac{1}{r}\frac{1+\sin\left(\frac{r}{r+s}\right)-\cos\left(\frac{r}{r+s}\right)}{\cos\left(\frac{r}{r+s}\right)} \;.
\ee
Inverting the Laplace transform formally we get
\be
Q_{4}\left(t\right)=\int_{\Gamma}\frac{ds}{2\pi i}\;e^{st}\;\frac{1}{r}\frac{1+\sin\left(\frac{r}{r+s}\right)-\cos\left(\frac{r}{r+s}\right)}{\cos\left(\frac{r}{r+s}\right)}\,,
\ee
where $\Gamma$ is a Bromwich integration contour in the complex plane.
Rescaling $s/r \to s$, we get $Q_{4}\left(t\right)=F_{4}\left(rt\right)$ with
\be
F_{4}\left(z\right)=\int_{\Gamma}\frac{ds}{2\pi i}\; e^{sz}\;\frac{1+\sin\left(\frac{1}{1+s}\right)-\cos\left(\frac{1}{1+s}\right)}{\cos\left(\frac{1}{1+s}\right)}\,.
\ee
This Bromwich integral can be evaluated by inspecting the poles of the integrand in the complex $s$-plane. The denominator, $\cos\left(\frac{1}{1+s}\right)$, vanishes on the negative real $s$ axis, when $\cos\left(\frac{1}{1+s}\right)=0$, i.e., at
\be
s_m=\frac{2}{\left(2m+1\right)\pi}-1 < 0,\quad m\in\mathbb{Z} \, .
\ee
Evaluating the numerator at $s=s_m$, we find
\be
1+\sin\left(\frac{1}{1+s_{m}}\right)-\cos\left(\frac{1}{1+s_{m}}\right)=1+\left(-1\right)^{m} \, .
\ee
\begin{figure}[h]
\includegraphics[width=0.45\textwidth]{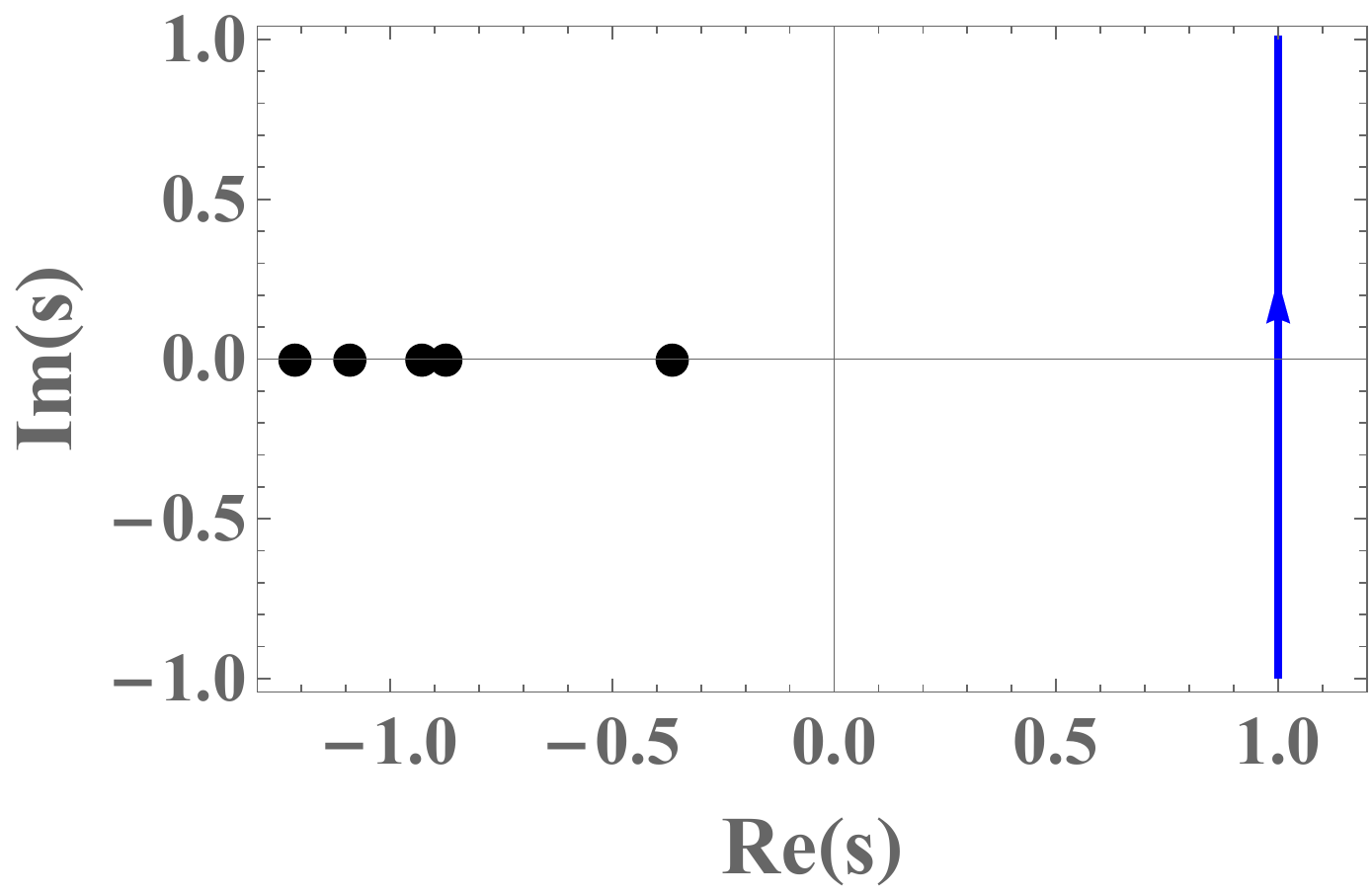}
\caption{One possible Bromwich integration contour in the complex $s$ plane, together with some of the poles, $s_m$ for $m = -4, -2, 0, 2, 4$.}
\label{FigBromwich}
\end{figure}
As we will see shortly, the integrand only has a pole for $m$ that is even (for $m$ that is odd, the numerator and denominator both vanish leading to a finite limit at $s \to s_m$, as we will see by showing that the residue vanishes for odd $m$).
The integration contour $\Gamma$ should be chosen such that it goes from minus complex infinity to plus complex infinity, and passes to the right of all of the poles in the complex $s$ plane.
One possibility for such a contour is plotted in Fig.~\ref{FigBromwich}, where the poles are also indicated.
\begin{figure*}[h]
\includegraphics[width=0.3\linewidth]{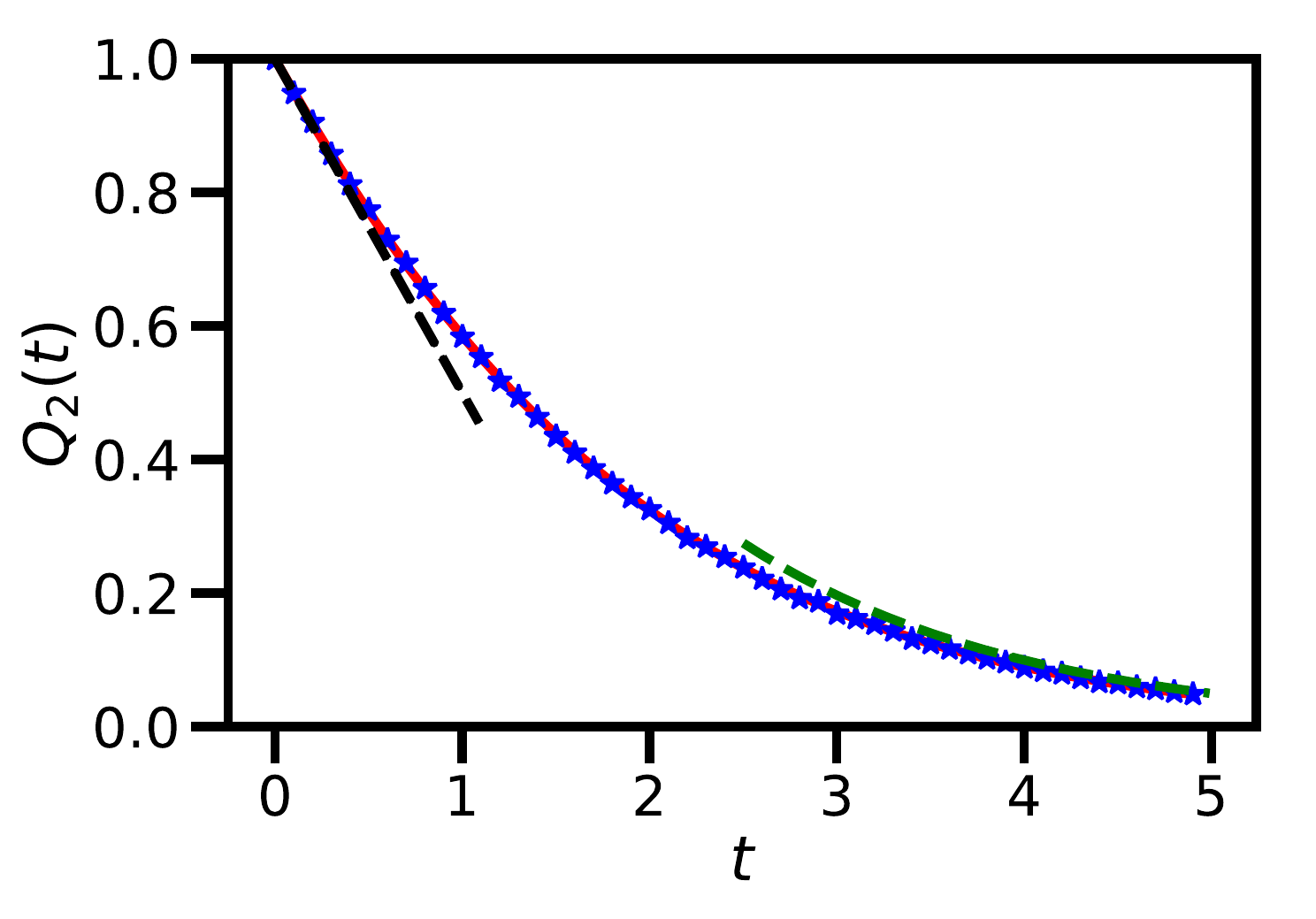} 
\includegraphics[width=0.3\linewidth]{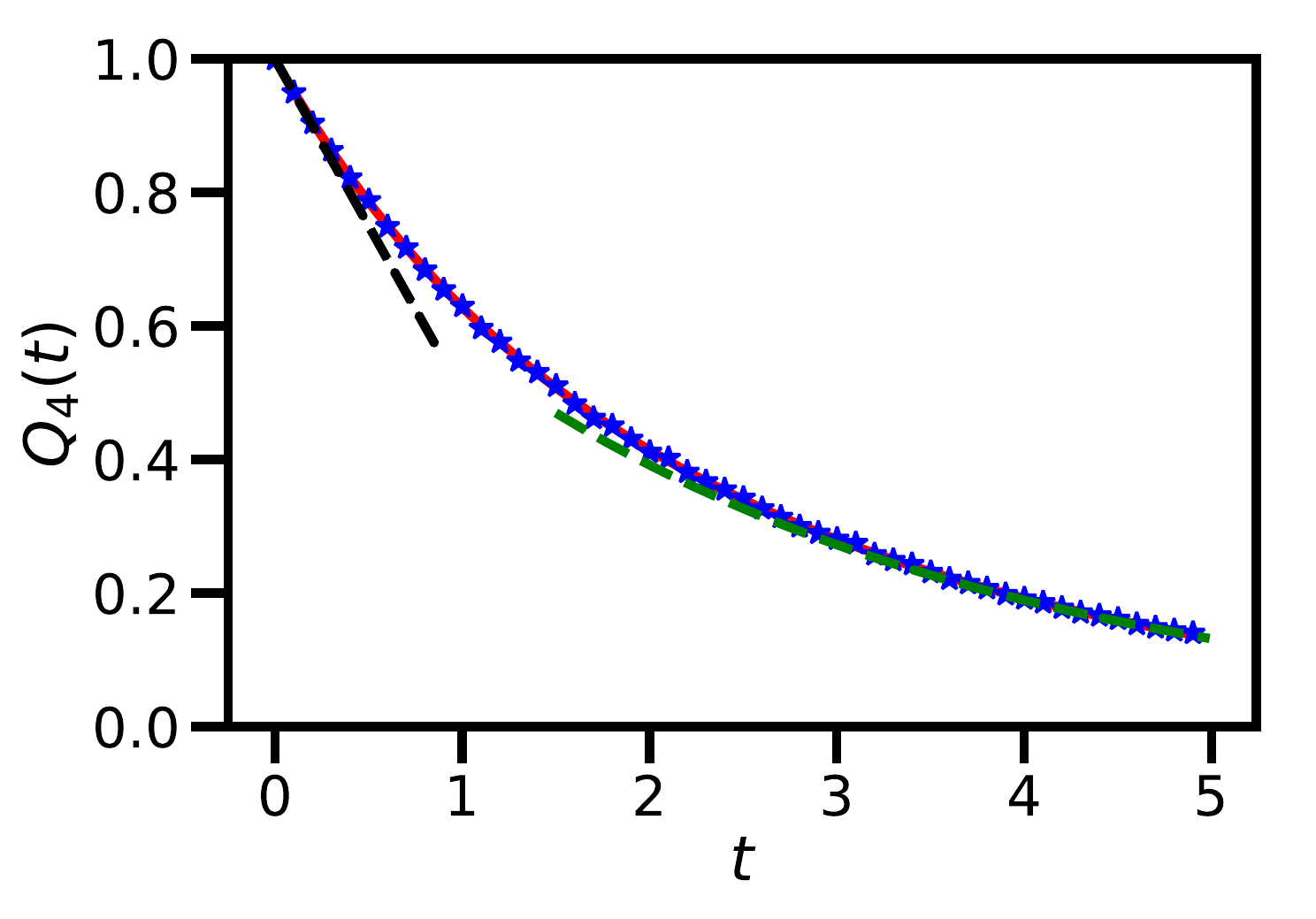}
\includegraphics[width=0.3\linewidth]{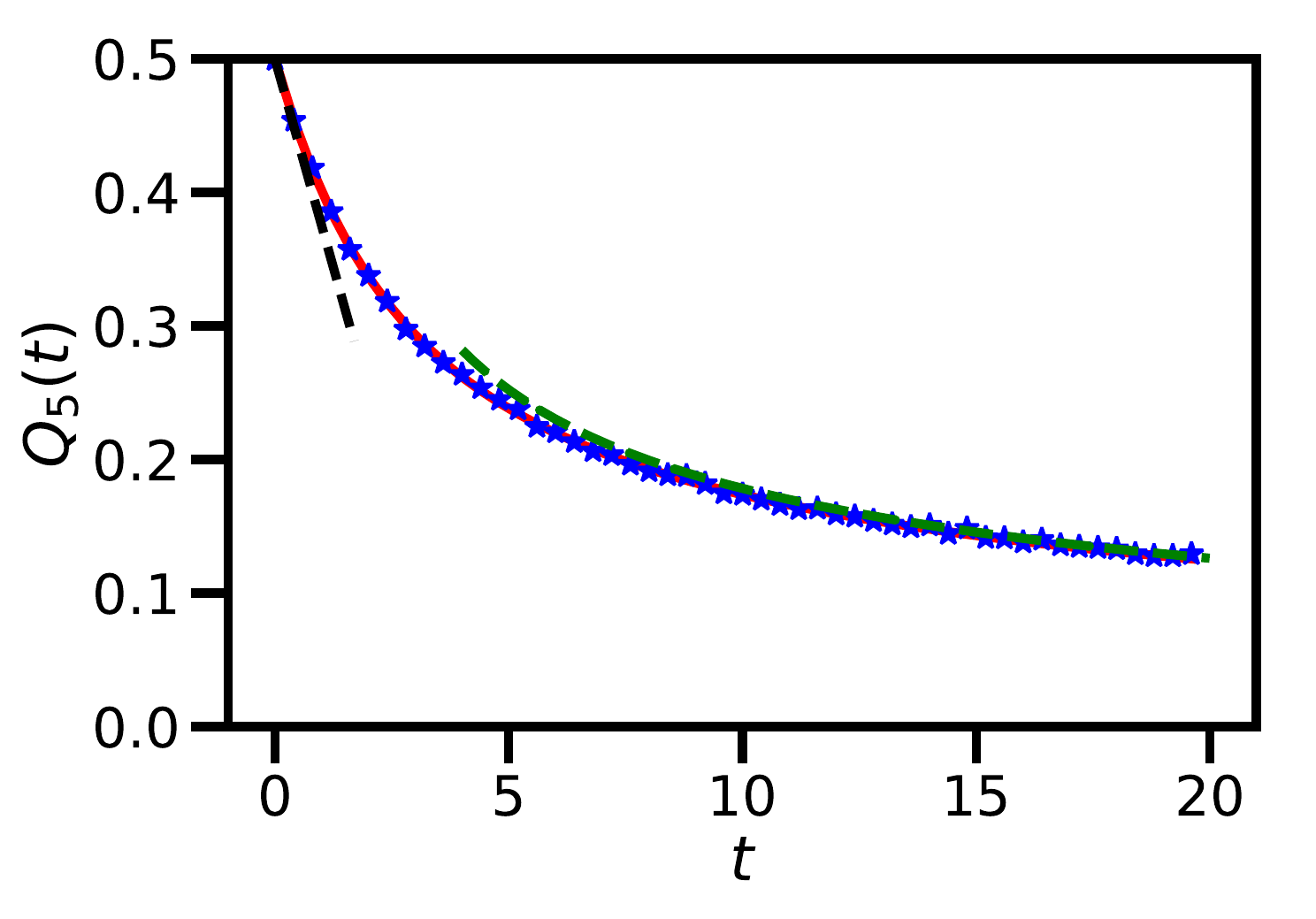}
\includegraphics[width=0.3\linewidth]{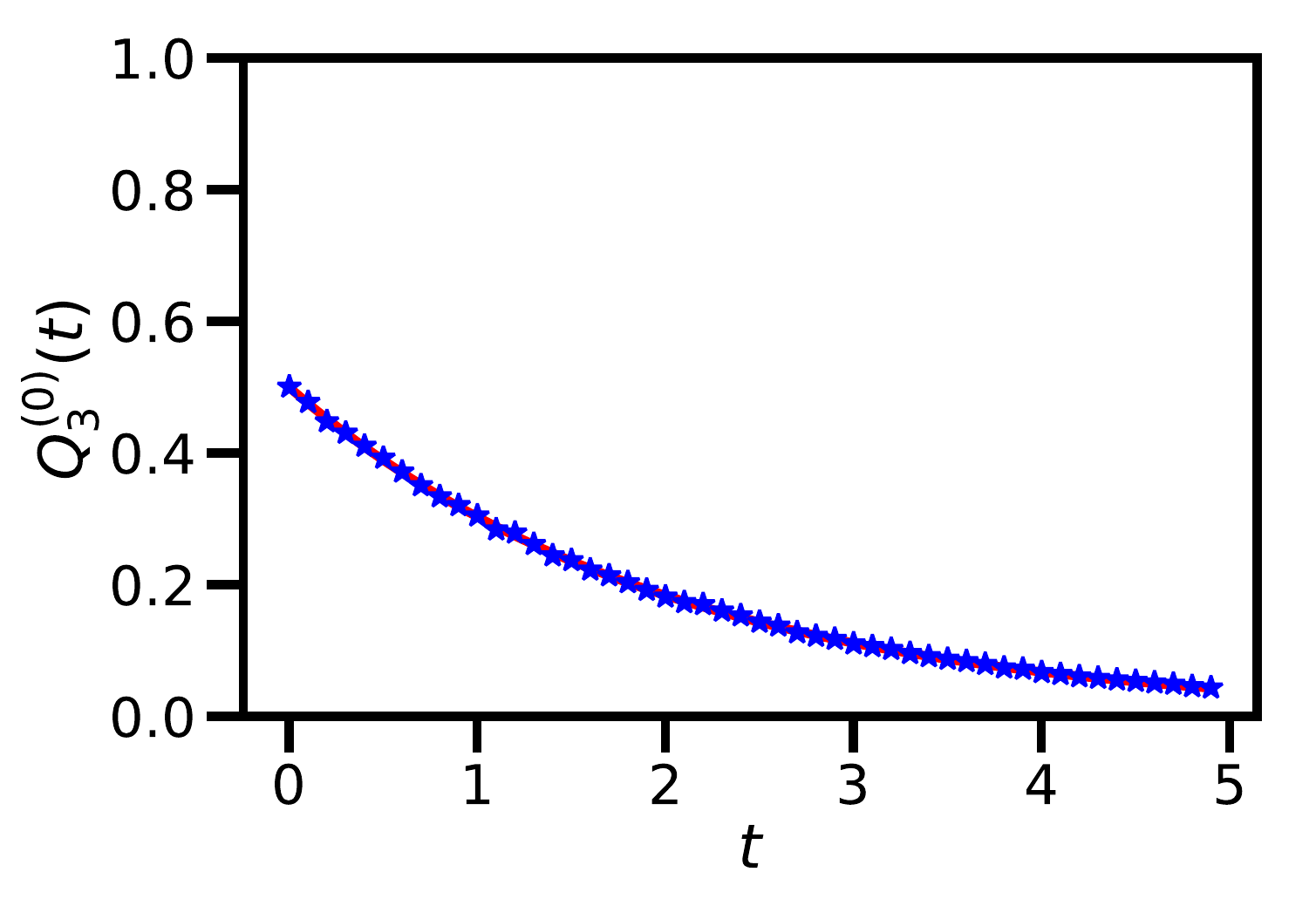}\hspace{7mm}
\includegraphics[width=0.3\linewidth]{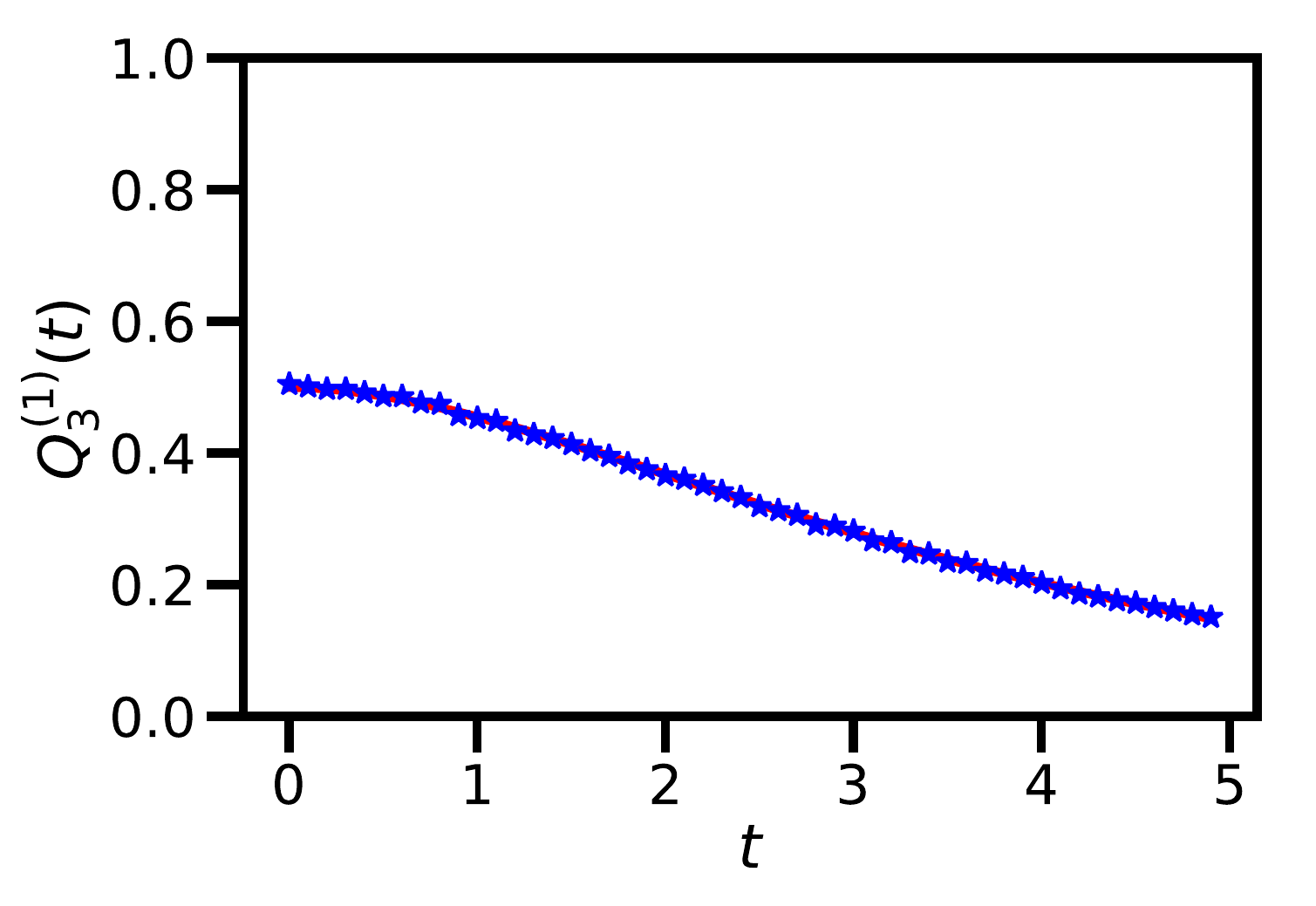}\hspace{7mm}
\includegraphics[width=0.3\linewidth]{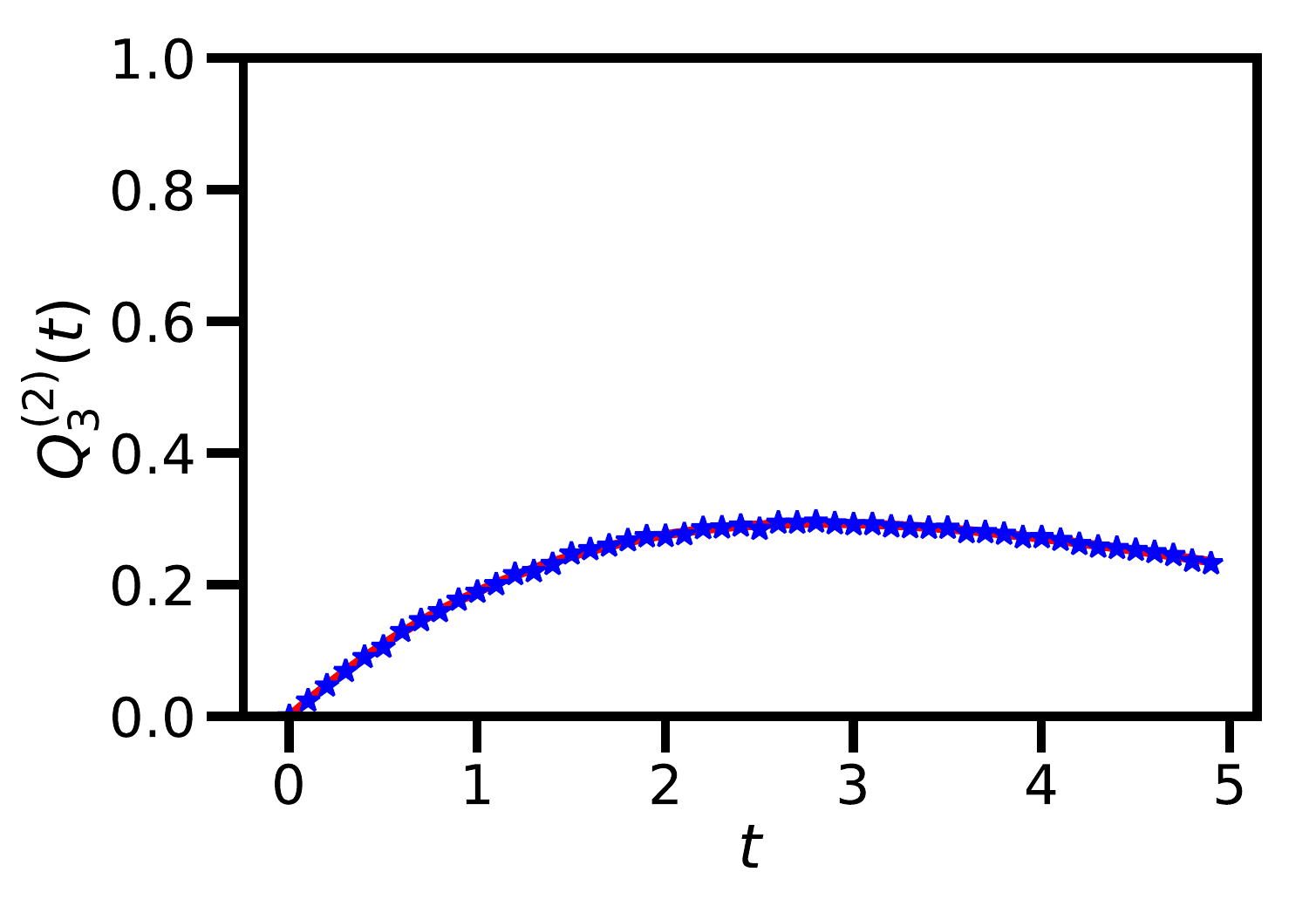}
\caption{Comparison of our predictions for $Q_i(t)$ with $i=2,3,4,5$ as a function of time $t$ (solid lines) with numerical results based on simulations of Brownian motion with resetting to the origin (symbols), showing perfect agreement.  Here we chose $r=1$. 
Dashed lines correspond to the asymptotic behaviors that are given in the main text.
For $Q^{(m)}_3(t)$, results are shown for $m=0,1,2$.}
\label{FigQs2345}
\end{figure*}
Now, $F_4(z)$ is found by summing over the residues at $s=s_m$ for $m \in \mathbb{Z}$:
\bea
&&F_{4}\left(z\right)= \sum_{m=-\infty}^{\infty}\text{Res}\left(s=s_{m}\right) \nn\\
&&=\sum_{m=-\infty}^{\infty}\lim_{s\to s_{m}}\left[\left(s-s_{m}\right)e^{sz}\frac{1+\sin\left(\frac{1}{1+s}\right)-\cos\left(\frac{1}{1+s}\right)}{\cos\left(\frac{1}{1+s}\right)}\right] \nn\\
&&= \sum_{m=-\infty}^{\infty}\left[e^{s_{m}z}\frac{1+\left(-1\right)^{m}}{\frac{\left(-1\right)^{m}}{\left(1+s_{m}\right)^{2}}}\right] \nn\\
&&=\frac{8}{\pi^{2}}\sum_{m=-\infty}^{\infty}\frac{1}{\left(4m+1\right)^{2}}e^{-\left(1-\frac{2}{\left(4m+1\right)\pi}\right)z} \, .
\eea
This reproduces the result in Eq.~\eqref{F4sol} in the main text.

\bigskip\bigskip

\noindent\textbf{Numerical verification of our results $Q_i(t)$ for $i=2,3,4,5$\\}

As a numerical check of our results $Q_i(t)$ for $i=2,3,4,5$, we performed simulations of reset Brownian motion with $U(z) = z$. The results, in Fig.~\ref{FigQs2345}, show perfect agreement with our analytical predictions given in the main text.

%
%




\end{document}